# Static interaction between electrically polarisable particles in vacuo

R. Coïsson
Dept. of physics and earth sciences
University of Parma, 43100 Parma, Italy
e-mail roberto.coisson@fis.unipr.it

**Abstract**
The static interaction of a point charge and a polarisable particle and between two polarisable particles is discussed in vacuo, and force and energy considerations are made. In particular a critical distance is shown (in principle) to appear in the two-dipole case, where the polarisation is self-sustained, and above which it disappears and below which it tends to explode. In the case of a polarisable particle with a nonzero charge interacting with a charge (of the same sign) there is a distance where repulsion and attraction are balanced.

**Introduction**
In order to clarify some basic aspects of classical electromagnetism, the interaction (in vacuo) of two point particles, either charge-charge, magnetic or electric dipole-dipole and charge-dipole are discussed in textbooks, and also in didactic articles, often discussing apparent paradoxes or showing relativistic aspects of classical electromagnetism (see [1-7]).
Along this line of thought, it may be interesting to consider also the case of point-like polarisable particles and their interactions with charges and between each other. In practice, in the literature similar considerations on elementary cases of molecular interactions are applied in order to describe (adding also quantum effects) interactions of ensembles of molecules in matter (see [8]).

**Charge and induced dipole**
In order to make a simple example, let us take an isotropic dipole (or one free to rotate), so the dipole is always aligned with the electric field, and we can consider the forces and movements in one dimension, so we will not use a vectorial notation.
A charge q is at the origin and a particle at position x has a dipole moment proportional to electric field:    $p = \alpha E$  (1)

The field produced by q at position x is   $E = \dfrac{q}{4\pi\epsilon_0} x^{-2}$  (2)

and the force on the induced dipole is   $F = p \dfrac{d}{dx} E = \alpha E \dfrac{dE}{dx} = \dfrac{-\alpha q^2}{8\pi^2 \epsilon_0^2 x^5}$  (3) that can be verified to be equal and opposite to the force on the charge q.
The work done against this force in order to move the particle (or the charge) from x to infinity:

$$W = -\int_x^\infty F \, dx = \dfrac{\alpha q^2}{8\pi^2 \epsilon_0^2} \int_x^\infty x^{-5} dx = \dfrac{\alpha q^2}{32\pi^2 \epsilon_0^2} x^{-4}$$ (4)

The work necessary to polarise the particle to the value it has at x is:

$$W_p = \dfrac{1}{2}\alpha E^2 = \dfrac{\alpha q^2}{32\pi^2 \epsilon_0^2} x^{-4}$$ (5)

If the dipole momenti is kept constant and we move it from x to infinity the work done is:

$$W_0 = \dfrac{\alpha q}{4\pi\epsilon_0} x^{-2} \int_x^\infty \dfrac{q}{2\pi\epsilon_0} x^{-3} dx = \dfrac{\alpha q^2}{16\pi^2 \epsilon_0^2} x^{-4}$$

(6)
We see that   $W_0 = 2 W_p = 2 W$  (7)

Then if we bring the particle to infinity keeping the dipole moment constant, we discharge the dipole and bring it back to position x (while it gets gradually charged), [or vice versa] the total work is zero, as it should be for energy conservation, as the system has come back to the initial condition.

**Two polarisable particles**
Let us consider two equal polarisable particles, with polarisability α. In this case there is no polarisation and then no interaction.
But imagine to use an external field $E_0$ to polarise them (in the same direction parallel to the distance vector, so they attract each other). Each of them, when reaching a polarisation p, produces an electric field on the other one:

$$E = \frac{p}{2\pi\epsilon_0} x^{-3} \tag{8}$$

If $\quad p = \alpha(E_0 + E_1) \tag{9}$

the added field produced at the other dipole is

$$E_1 = \frac{\frac{\alpha}{2\pi\epsilon_0 x^3} E_0}{1 - \frac{\alpha}{2\pi\epsilon_0 x^3}} \tag{10}$$

When $\quad x = \left(\frac{\alpha}{2\pi\epsilon_0}\right)^{1/3} \tag{11}$

the situation sustains itself: no external force is necessary to polarise them ($E_0=0$); only a force is necessary to keep them apart, as they are attracting each other.

In this case the polarisation is self-sustained, and any value of p (and then E) is in (indifferent) equilibrium. But the equilibrium is unstable with respect to the distance x: at a greater distance the polarisations disappear, and at shorter distances the dipoles explode (or reach a saturation value due to a possible nonlinearity). This effect. however, might be of little practical importance, as for a molecule the critical distance would be of the order of a nanometre or less, and then the point-like approximation would no longer be valid.

If the two polarisabilities are different, say α and β, for the critical distance $\alpha \rightarrow (\alpha\beta)^{1/2}$ and the ratio of the two equilibrium dipoles is $\quad \frac{p_1}{p_2} = \left(\frac{\alpha}{\beta}\right)^{1/2} \tag{12}$

The energy of this system (in equilibrium) can be calculated (again for two equal dipoles) by imaging to discharge one of the two dipoles while keeping the other one fixed (in order to keep it fixed we need a force, but no work) and then discharge the other one: for the first operation we need to do work, as we have to counteract the effect of the constant field E, and for the second one we get the energy corresponding to its polarisation (in a zero field, as the other dipole is now zero): as a consequence $\quad W = \frac{p^2}{2\alpha} - \frac{p^2}{2\alpha} = 0 \tag{13}$

But we may also keep both dipoles fixed and move one of them to infinity: the work necessary is

$$\int_x^\infty F\,dx = \int_x^\infty p\frac{dE}{dx}dx = \frac{p^2}{2\pi\epsilon_0 x^3} = \frac{p^2}{\alpha} \tag{14}$$

.and the same energy is obtained from discharging the dipoles, then again the balance is zero. Also we can remark that in this case of indifferent equilibrium, no work is necessary in order to modify the polarisation, from zero to any value.

In the more general case of a polarisation with a finite value at zero field: $\quad p = p_0 + \alpha E$ we get

$$E = \frac{p_0}{2\pi\epsilon_0 x^3}\left(1 - \frac{\alpha}{2\pi\epsilon_0 x^3}\right)^{-1} \tag{15}$$

the dipoles explode when $\quad x^3 \leq \frac{\alpha}{2\pi\epsilon_0}$ , and for higher values of x the polarisation is increased

(with respect to the permanent dipole) by a factor $\dfrac{p}{p_0} = 1 + \dfrac{\alpha}{2\pi\epsilon_0 x^3}\left(1 - \dfrac{\alpha}{2\pi\epsilon_0 x^3}\right)^{-1}$ (16)

**Polarisable particle with charge**

Let us now consider a particle which has a charge q and is at the same time polarisable, .
If this is (at position x) in the field produced by a fixed charge Q of the same sign (at the origin), it is subject to a force

$$F = \dfrac{qQ}{4\pi\epsilon_0 x^2} - \dfrac{pQ}{2\pi\epsilon_0 x^3} = \dfrac{qQ}{4\pi\epsilon_0 x^2} - \dfrac{\alpha Q^2}{8\pi^2\epsilon_0^2 x^5} \tag{17}$$

The repulsive force between the two charges is balanced by the attractive force on the induced dipole when

$$x^3 = \dfrac{\alpha Q}{2\pi\epsilon_0 q} \quad : \tag{18}$$

at smaller distances the force is attractive, so particles tend to condensate.

If both charged particles are polarisable, The result is slightly more complicated. Let us consider for simplicity two identical particles of charge q and polarisability α.
At the position of each particle we have

$$E = \dfrac{q}{4\pi\epsilon_0 x^2} - \dfrac{\alpha E}{2\pi\epsilon_0 x^3} \quad \text{then} \quad E = \dfrac{q}{4\pi\epsilon_0 x^2\left(1 + \dfrac{\alpha}{2\pi\epsilon_0 x^3}\right)} \tag{19}$$

The force is $F = qE + \alpha E \dfrac{dE}{dx} = \dfrac{\beta q^2}{x^2} - \dfrac{2\beta^2 \alpha q^2}{x^5} + \dfrac{2\beta^3 \alpha^2 q^2}{x^8}$ (20)

where $\beta = \dfrac{1}{4\pi\epsilon_0\left(1 + \dfrac{\alpha}{2\pi\epsilon_0 x^3}\right)}$ (20')

there are two repulsive forces: charge-charge and dipole-dipole (as the same poles face each other) and two attractive forces: charge-dipole and dipole-charge (the latter two are equal if the charges and polarisabilities are equal), and there is an interval of x where the force is attractive.

**Acknowledgements**
The author thanks G. Asti for many discussions and comments, D. J. Griffiths for a useful exchange of e-mails and E. G. Bessonov for detailed comments and help in improving the text.